\documentclass[twocolumn,showpacs]{revtex4}

\usepackage{graphicx}%
\usepackage{dcolumn}
\usepackage{amsmath}

\makeatletter
\def\btt#1{\texttt{\@backslashchar#1}}%
\DeclareRobustCommand\bblash{\btt{\@backslashchar}}%
\makeatother

\begin{document}

\preprint{HEP/123-qed}

\title[Short Title]{
Origin for the enhanced copper spin echo decay rate in the pseudogap regime\\ 
of the multilayer high-$T_{\rm c}$ cuprates
}
 
\author{Atsushi Goto$^{1,2}$}
\author{W. G. Clark$^1$}
\author{Patrik Vonlanthen$^1$}
\author{Kenji B. Tanaka$^1$}
\author{Tadashi Shimizu$^2$}
\author{\\Kenjiro Hashi$^2$}
\author{P. V. P. S. S. Sastry$^{3,4}$}
\author{Justin Schwartz$^{3,4}$}

\affiliation{
$^{1}$Department of Physics and Astronomy, 
University of California, Los Angeles, CA, 90095-1547}
\affiliation{
$^{2}$National Institute for Materials Science, 
3-13, Sakura, Tsukuba, Ibaraki, 305-0003, Japan}
\affiliation{
$^{3}$National High Magnetic Field Laboratory, 
1800 E. Paul Dirac Drive, Tallahassee, Florida 32310}
\affiliation{
$^{4}$Department of Mechanical Engineering, 
FAMU-FSU College of Engineering, Tallahassee, Florida 32306}

\date{\today}% It is always \today, today, but you may specify any date with \date.
\vspace{1cm}
\begin{abstract}
We report measurements of the anisotropy of the spin echo decay 
for the inner layer Cu site of the triple layer cuprate, 
Hg$_{0.8}$Re$_{0.2}$Ba$_2$Ca$_2$Cu$_3$O$_8$ ($T_{\rm c}$=126 K) 
in the pseudogap $T$ regime below $T_{\rm pg} \sim $170 K
and the corresponding analysis for their interpretation.
As the field alignment is varied, 
the shape of the decay curve changes 
from Gaussian ($H_0 \parallel$ c) to single exponential ($H_0 \perp$ c).
The latter characterizes the decay caused by the fluctuations 
of adjacent Cu nuclear spins caused by their interactions with electron spins.
The angular dependence of the second moment
($T_{\rm 2M}^{-2}$ $\equiv$ $<\Delta \omega^2>$) 
deduced from the decay curves indicates that
$T_{\rm 2M}^{-2}$ for $H_0 \parallel$ c, 
which is identical to $T_{\rm 2G}^{-2}$ ($T_{\rm 2G}$ is the Gaussian component), 
is substantially enhanced, 
as seen in the pseudogap regime of the bilayer systems.
Comparison of $T_{\rm 2M}^{-2}$ between $H_0 \parallel$ c and $H_0 \perp$ c 
indicates that this enhancement is caused by electron spin correlations 
between the inner and the outer CuO$_2$ layers.
These results provide the answer to the long-standing controversy
regarding the opposite $T$ dependences of $(T_1T)^{-1}$ and $T_{\rm 2G}^{-2}$ 
in the pseudogap regime of bi- and trilayer systems.
\end{abstract}

\maketitle

The pseudogap phenomenon has been one of the central issues 
in understanding the anomalous normal states of high-$T_{\rm c}$ cuprates.
Since the discovery of the pseudogap \cite{yasuoka89}, NMR has continued to be a basic tool for its investigation. Temperature and doping dependences of the spin-lattice relaxation times ($T_1$) 
and the Knight shifts ($K_{\rm s}$) of the Cu sites 
have served as crucial tests for the theories describing the pseudogap.
A controversy, however, has arisen on the interpretation of 
the Gaussian component of a spin-spin relaxation time ($T_{\rm 2G}$).
In systems such as YBa$_2$Cu$_3$O$_{6.6}$ and YBa$_2$Cu$_4$O$_8$, 
$T_{\rm 2G}^{-2}$ continues to grow as $T$ is lowered towards $T_{\rm c}$,
whereas $(T_1T)^{-1}$ decreases in the pseudogap regime \cite{itoh92as,takigawa92}. If the same dynamical susceptibility $\chi({\bf q}, \omega)$ is responsible for both relaxation rates, an anomalous enhancement is expected at the high frequency part of Im$\chi({\bf q}, \omega)$ in the pseudogap regime. 
This rather peculiar conclusion has puzzled theorists, 
and other explanations have been sought.

A key to resolve the problem is that the phenomenon is observed only in multilayer systems.
One proposed mechanism is spin correlations between adjacent layers 
\cite{kishine96,kishine97,goto98},
which had been observed in spin echo double resonance (SEDOR) experiments
\cite{stern95,suter99,statt97}. 
They are expected to play essential roles in the systematic increase of $T_{\rm c}$ 
with increasing the number of layers
because they can provide attractive forces between layers and stabilize the superconductivity
\cite{ubbens94}.
In the bilayer systems, the two CuO$_2$ layers in a unit cell are equivalent,
so that the Cu nuclei on one of the layers behave as like-spins to the others 
in the echo decay process,
and contribute to $T_{\rm 2G}^{-2}$ through the interlayer spin correlations.
Unfortunately, since the contribution from the interlayer couping is indistinguishable 
from its intralayer counterpart for the identical layers,
it is difficult to extract the effect experimentally in such systems.

In order to identify experimentally the interlayer effects on $T_{\rm 2G}$, 
we have utilized the trilayer cuprate 
Hg$_{0.8}$Re$_{0.2}$Ba$_2$Ca$_2$Cu$_3$O$_8$ ($T_{\rm c}=126$ K) 
with a pseudogap ($T_{\rm pg} \sim 170$ K),
where one inner and two outer CuO$_2$ layers are crystallographically inequivalent.
This enables us to separate the interlayer effects from the total decay rates.
In this letter, we report the measurements of the angular dependence 
of the second moment in the inner Cu site,
which identifies the role of interlayer correlations in the echo decay process
in the pseudogap regime of a multilayer system.
Our analysis of the results shows that 
the different behavior of $(T_1T)^{-1}$ and $T_{2\rm G}^{-2}$ is caused by interlayer spin correlations.

The powder sample was prepared using the method of Ref. \onlinecite{sastry98as}
and magnetically oriented along the c-axis.
\begin{figure}[t]
\includegraphics[scale=0.43]{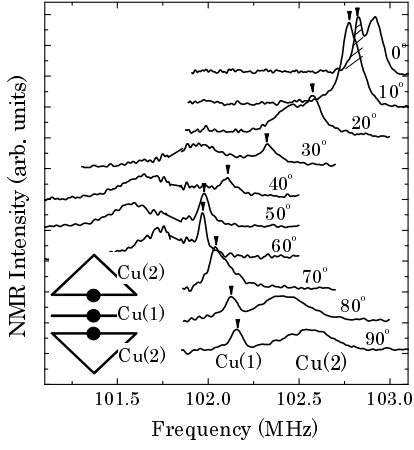}% Here is how to import EPS art
\caption{Absorption spectra for the central transition of $^{63}$Cu at 135 K
as a function of $\theta$.
The triangles show the part of the spectra used for the $T_2$ measurements.
Inset: Schematic view of the CuO$_2$ layers.}
\label{spectrum}
\end{figure}
Figure \ref{spectrum} shows the angular dependence of the frequency spectrum 
for the $^{63}$Cu central transition at 135 K.
The angle $\theta$ is that between $H_0$ and the c-axis.
Each curve was obtained by adding the real part of the FT spectra of the half echo
measured at a few different frequencies \cite{clark95}.
The relatively narrow and wide lines are assigned to Cu(1) and Cu(2) sites in the inner and outer layers, respectively
\cite{goto99a}.
We confirmed that $K_{\rm s}$ and the quadrupole frequency ($\nu_{\rm Q}$) are consistent 
with previous results \cite{magishi96s,julien96s}.
The triangles in Fig. \ref{spectrum} indicate the positions 
at which echo decays were measured.  
Since the Cu(1) and Cu(2) lines overlap at 10$^{\circ}$ and 70$^{o}$,
both lines contribute to the measured decay.

\begin{figure}[t]
\includegraphics[scale=0.34]{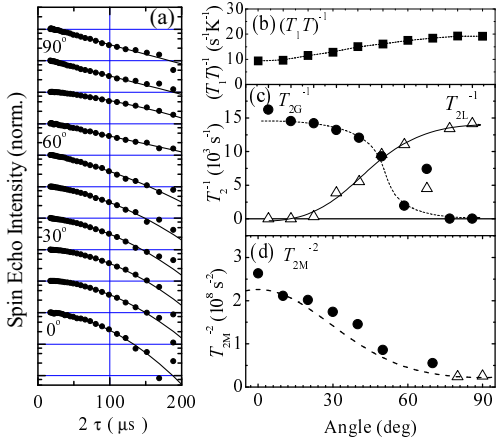}
\caption{Angular dependences of (a) the echo decay curve, (b) $(T_1T)^{-1}$, 
(c) $T_{\rm 2G}^{-1}$ ($\bullet$) and $T_{\rm 2L}^{-1}$ ($\triangle$) in Eq. (\ref{GL-fit}) and 
(d) $T_{\rm 2M}^{-2}$ deduced from $T_{\rm 2G}^{-2}$ ($\bullet$) and $T_{\rm 2L}^{-2}$ ($\triangle$) at 135 K.
The dashed curve in (d) is given by Eq. (\ref{T2M-aniso}).}
\label{anisotropy}
\end{figure}

The angular dependence of the echo decay curves at 135 K is shown in Fig. \ref{anisotropy}(a),
where the Redfield contribution from $T_1$ has been removed 
by dividing the measured data by $\exp(-2\tau/T_{1R})$\cite{redfield57}
with $T_{1R}$ obtained as follows \cite{walstedt95}.
The general form of $T_{\rm 1R}^{-1}$ is given by 
\begin{equation}
(T_{1R})_z^{-1}=\{I(I+1)-1/4\}(W_x+W_y)+W_z,
\label{T1R}
\end{equation}
where $W_{\gamma}$ ($\gamma$=x, y, z) is from the spin fluctuations 
in the $\gamma$ direction and $z$ is the quantization axis ($\parallel H_0$).
In the same notation, $(T_1)_z^{-1}$ is given by
$ (T_1)_z^{-1}=W_x+W_y.$
Hence, for an arbitrary $\theta$, 
\begin{eqnarray}
  [T_{1R}(\theta)]^{-1}&=&\{I(I+1)-1/4\}[T_1(\theta)]^{-1} \nonumber\\
  & & + [T_1(90^{\circ}-\theta)]^{-1}-0.5[T_1(0^{\circ})]^{-1},
\end{eqnarray}
where the relation $W_a=W_b$ is used (The subscripts $a,b$ correspond to 
the crystalline axes). 
From the anisotropy of $(T_1T)^{-1}$ in Fig. \ref{anisotropy}(b),
$[T_{1R}(\theta)]^{-1}$ is calculated. 

Figure \ref{anisotropy}(a) shows that the shape of the decay curve changes from 
Gaussian at 0$^{\circ}$ to single exponential (Lorentzian spectrum) at 90$^{\circ}$.
These data are fitted by the function,
\begin{equation}
 \frac{M(2 \tau)}{e^{-2\tau/T_{1R}}}=M_0 \exp \left [-\frac{2\tau}{T_{\rm 2L}}-\frac{1}{2}\left (\frac{2\tau}{T_{\rm 2G}} \right )^2 \right ]
\label{GL-fit}
\end{equation}
with $M_0$, $T_{\rm 2L}$ and $T_{\rm 2G}$ as free parameters.
The angular dependences of $T_{\rm 2L}^{-1}$ and $T_{\rm 2G}^{-1}$ thus obtained 
are shown in Fig. \ref{anisotropy}(c).
For $0^{\circ} \sim 20^{\circ}$ , $ T_{\rm 2L}^{-1}$ is nearly zero 
while $T_{\rm 2G}^{-1}$ is zero at $80^{o}$ and 90$^{o}$.
In between, the decay curve crosses over between these two extremes.
This change is a result of the NMR line narrowing caused by the following mechanism.
At finite $T$ the effective interaction between adjacent nuclear spins is reduced because of rapid spin flips driven by the hyperfine interaction with the electrons, 
which averages out the nuclear fields at adjacent nuclear sites.
As a result, the central part of the NMR line is narrowed 
and the spectrum approaches a Lorentzian rather than a Gaussian shape
\cite{abragam61,walstedt95}.
The importance of the effect depends on the ratio 
between the two time scales $T_1$ and $T_{\rm 2M} \equiv \langle \Delta \omega^2 \rangle^{-1/2}$, where $\langle \Delta \omega^2 \rangle$ is the homogeneous second moment of the NMR absorption.
The former and the latter characterize the time scales of the nuclear spin fluctuations
and the echo decays, respectively.

There are two limiting cases where analytical forms for the decays are known.
One is the static limit ($T_1/T_{\rm 2M} \gg 1$), 
where nuclear spins do not change their states between pulses or a pulse and an echo 
because of the relatively long $T_1$.
Consequently, the contributions from unlike-spins are canceled out 
at the time of the echo and only like-spins contribute to the echo decay. 
The decay in this case is described by a Gaussian \cite{itoh92as},
\begin{equation}
 \frac{M(2 \tau)}{e^{-2 \tau/T_{1R}}}= M_0 \exp \left [-\frac{1}{2}\left (\frac{2 \tau}{T_{\rm 2M}}\right )^2 \right ] f(2\tau),
\label{static}
\end{equation}
where $f(2\tau)$ is the correction for the narrowing effect \cite{curro97} 
and in the static limit, $f(2 \tau) \equiv 1$. 
The rate $T_{\rm 2M}^{-1}$ in this case is usually referred to as the Gaussian rate, $T_{\rm 2G}^{-1}$.

The second is the narrowed limit ($T_1/T_{\rm 2M} \le 1$), where nuclei are fluctuating during the echo sequence.
The decay is characterized by a single exponential (Lorentzian spectrum) \cite{abragam61,walstedt95},
\begin{equation}
 \frac{M(2 \tau)}{e^{-2 \tau/T_{1R}}} = M_0 \exp \left [
-2 \tau \left (\frac{^{63}T_1}{^{63}T_{\rm 2M}\smallskip^2}+\frac{^{65}T_1}{^{65}T_{\rm 2M}\smallskip^2} \right ) \right ],
\label{narrowing}
\end{equation}
where $^{\alpha}T_{\rm 2M}^{-2} \equiv\hspace{-0.2mm}^{\alpha}\langle \Delta \omega^2 \rangle$ 
is the contribution from the $\alpha$-nuclei ($\alpha$=63, 65) to the second moment of $^{63}$Cu,
and $^{\alpha}T_1$ is the spin-lattice relaxation time in the $\alpha$-site. 
Note that $^{63}T_{\rm 2M}$ corresponds to $T_{\rm 2G}$ in the static limit.
Here, $^{65}$Cu also contributes to the $^{63}$Cu decay 
because they lose their memories of the initial states during the echo sequence 
due to the fluctuation effect,
so that their contributions are not canceled out at the time of the $^{63}$Cu echo. 

In the present case, the transition from the static to the narrowing regime
is caused by the large anisotropies of $T_1^{-1}$ and $T_{\rm 2M}^{-1}$.
As will be shown later, $T_{\rm 2M}^{-1}$ decreases by 3.3 from $\theta=0^{\circ}$ to 90$^{\circ}$,
while $T_1^{-1}$ increases by 2.1. 
Also, there is a qualitative correspondence between
the transition of the decay curve in Fig. \ref{anisotropy}(a) and
the simulations by Walstedt et al. for the various values of $T_1/T_{\rm 2M}$
(Fig. 4 of Ref \onlinecite{walstedt95}).
The angular dependence of $T_{\rm 2M}^{-2}$ 
deduced from either $T_{\rm 2G}$ or $T_{\rm 2L}$ in Fig. \ref{anisotropy}(c)
is shown in Fig. \ref{anisotropy}(d),
where $T_{\rm 2M}=T_{\rm 2G}$
while Eq. (\ref{narrowing}) is used to obtain $^{63}T_{\rm 2M}$ from $T_{\rm 2L}$
along with the relations,
\begin{eqnarray}
  (^{63}\gamma)^2 \cdot ^{63}T_1 & = & (^{65}\gamma)^2 \cdot ^{65}T_1,\\
^{65}c \cdot (^{65}\gamma_n \cdot ^{65}T_{\rm 2M})^2  & = & ^{63}c \cdot (^{63}\gamma_n \cdot ^{63}T_{\rm 2M})^2,
\label{eqn:T1-relation}
\end{eqnarray} 
where $^\alpha c$ is the natural abundance for the isotope $\alpha$.

The angular dependence of $T_{\rm 2M}^{-2}$ in Fig. \ref{anisotropy}(d) is consistent with
that of the hyperfine coupling constant.
In the detuned limit, 
the flip-flop term ($I_i^+I_j^-$) in the nuclear Hamiltonian is ineffective 
because of the mismatch in the Zeeman energies between adjacent nuclei. 
Hence, $T_{\rm 2M}^{-2}$ is given only by the z-component term ($I_i^zI_j^z$), so that
%\begin{equation}
$ [T_{\rm 2M}(\theta)]^{-2} \propto \{\chi({\bf Q})\}^2 \cdot \sum_q \{F_{\bf q}(\theta)\}^2$,
%\label{t2-expression}
%\end{equation}
where $F_{\bf q}(\theta)$ is the form factor when $H_0$ is in the $\theta$ direction
\cite{takigawa92}.
Here, we assume that the q-dependence of $\chi({\bf q})$ around ${\bf Q}\equiv (\pi, \pi)$ 
is weaker than that of $F_{\bf q}$,
so that $\chi({\bf q})$ is represented by $\chi({\bf Q})$ and taken out of the q-summation.
We further assume that $\sum_q \{F_{\bf q}(\theta)\}^2$ is proportional to 
$\{F_{\bf Q}(\theta)\}^2$ at each $\theta$.
Since the $\theta$ dependence of $F_{\bf Q}(\theta)$ 
is $\{A(3\cos^2{\theta}-1)+B\}^2$
where $A$ and $B$ are constants,
the anisotropy of $T_{\rm 2M}^{-2}$
% given by Eq. (\ref{t2-expression}) 
is given by
\begin{equation}
  [T_{\rm 2M}(\theta)]^{-2}/[T_{\rm 2M}(90^{o})]^{-2}=(\sin^2{\theta}+\xi \cos^2{\theta})^4,
\label{T2M-aniso}
\end{equation}
where 
\begin{equation}
 \xi \equiv \left [ \frac{\sum_q \{F_{\bf q}(0^{\circ})\}^2}{\sum_q \{F_{\bf q}(90^{\circ})\}^2} \right ]^{1/4} 
\approx \left [\frac{F_{\bf Q}(0^{\circ})}{F_{\bf Q}(90^{\circ})} \right ]^{1/2}.
\end{equation} 
The value of $\xi$ can be estimated from the anisotropy of $T_1^{-1}$. 
Since $\chi({\bf Q}) \gg \chi(0)$ in this system,
$(T_1)_z^{-1} \propto \{F_x({\bf Q})+F_y({\bf Q})\}$ \cite{magishi96s}.
Hence, $\xi$ is given by,
\begin{equation}
\xi \approx \left [\frac{F_{\bf Q}(0^{\circ})}{F_{\bf Q}(90^{\circ})} \right ]^{1/2} 
\approx \left [2\frac{(T_1(90^{\circ}))^{-1}}{(T_1(0^{\circ}))^{-1}}-1 \right ]^{1/2}.
\end{equation}
From Fig. \ref{anisotropy}(b), $(T_1(90^{\circ}))^{-1}/(T_1(0^{\circ})^{-1}$ = 2.1, so that $\xi$=1.79. 
The dashed curve in Fig. \ref{anisotropy}(d) shows
the anisotropy of $T_{\rm 2M}^{-2}$ obtained from Eq. (\ref{T2M-aniso}) 
with $[T_{\rm 2M}(90^{\circ})]^{-2}$ as a single adjustable parameter.
It has the same tendency as the angular dependence of $T_{\rm 2G}^{-2}$
in spite of some assumptions.

The analysis at 0$^{\circ}$, 10$^{\circ}$ and 70$^{\circ}$ is not straightforward
because of the overlap with the Cu(2) line. 
Since $T_{\rm 2M}^{-1}$ at Cu(2) is expected to be smaller than that of Cu(1) 
by a factor of 0.75 \cite{magishi96s}, 
the overlapped Cu(2) line reduces $T_{\rm 2M}^{-1}$. 
$T_{2M}^{-2}$, however, is not reduced at 10$^{\circ}$, and is significantly enhanced at 0$^{\circ}$.
At 70$^{\circ}$, the shape of the decay curve itself is quite different 
from those at 60$^{\circ}$ and 80$^{\circ}$.
Below, we show that these features can be attributed 
to the interlayer spin correlations,
which have the effect of enhancing $T_{\rm 2M}^{-1}$ \cite{kishine96,goto98}. 

Consider the echo decay process at $\theta=0^{\circ}$.
As seen in Fig. \ref{spectrum}, the Cu(1) and Cu(2) lines
are situated close to each other, so that
not only Cu(1) but also a part of Cu(2) nuclei are excited,
which also act as like-spins for the Cu(1) nuclei in the echo decay process.
On the other hand, the intensity of the echo is obtained 
by integrating only the Cu(1) part of the FT spectrum of the echo 
(shaded part of the spectrum in Fig. \ref{spectrum}), 
so that only the Cu(1) nuclei contribute 
to the intensity of the decay curve in Fig. \ref{anisotropy}(a).
This is the same situation as that in the SEDOR experiment 
where the $\pi$-pulses for like- and unlike-spins are applied simultaneously.
\begin{figure}[t]
\includegraphics[scale=0.35]{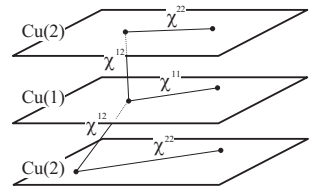}% Here is how to import EPS art
\caption{Schematic view of the intra ($i=j$) and interlayer ($i \neq j$) 
spin susceptibilities ($\chi^{ij}({\bf q})$) in the trilayer system.}
\label{interlayer}
\end{figure}
Since all the like-spins contribute to the Gaussian decay in the static limit,
$[T_{\rm 2M}(0^{\circ})]^{-2}$ is given by \cite{goto98},
\begin{equation}
[T_{\rm 2M}(0^{\circ})]^{-2} \propto \sum_q [F_{\bf q}(0^{\circ}) \{\chi^{11}({\bf q})+4 \epsilon \chi^{12}({\bf q})\}]^2,
\label{tri-t2-in}
\end{equation}
where, $\chi^{11}$ and $\chi^{12}$ are the intra- and interlayer spin susceptibilities 
associated with the auto- and cross-correlations within or between layers
indicated in Fig. \ref{interlayer}.
The second term in Eq. (\ref{tri-t2-in}) corresponds to the contribution 
due to the interlayer correlations.
The ratio of the excited Cu(2) nuclei ($\epsilon$) is estimated to be about 0.6.

At the angles where the two Cu lines are separated from each other,
the second term in Eq. (\ref{tri-t2-in}) does not appear in the echo decay process, 
whereas at 10$^{\circ}$ and 70$^{\circ}$, both the Cu(1) and Cu(2) nuclei are excited and observed, 
so that the second term in Eq. (\ref{tri-t2-in}) also appears, 
which enhances $T_{\rm 2M}^{-2}$.
This enhancement increases $T_1/T_{\rm 2M}$,
and brings the situation at 70$^{\circ}$ closer to the static limit, 
resulting in the appearance of the Gaussian component in the decay curve. 
At 10$^{\circ}$, a cancelation may occur between the reduction due to the
overlapped Cu(2) line and the enhancement due to the interlayer spin correlations.

\begin{figure}[t]
\includegraphics[scale=0.42]{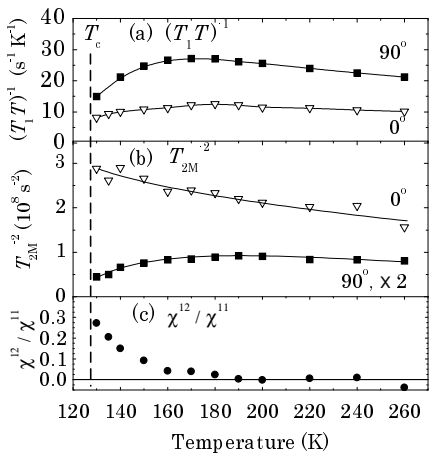}% Here is how to import EPS art
\caption{$T$ dependences of (a) $(T_1T)^{-1}$ and (b) $T_{\rm 2M}^{-2}$ 
at 0$^{\circ}$ and $2T_{\rm 2M}^{-2}$ at $90^{\circ}$. (c) $\chi^{12}({\bf Q})/\chi^{11}({\bf Q})$ defined by Eq. (\ref{rchi}).}
\label{temperature}
\end{figure}
Figure \ref{temperature}(b) shows the $T$ dependences 
of $T_{\rm 2M}^{-2}$ at 0$^{o}$ and 90$^{\circ}$,
which are quite different from each other; i.e.,
while $[T_{\rm 2M}(90^{\circ})]^{-2}$ starts to decrease at $T_{\rm pg}$ 
as does $(T_1T)^{-1}$ shown in Fig. \ref{temperature}(a), 
$[T_{\rm 2M}(0^{\circ})]^{-2}$ continues to grow down to $T_{\rm c}$.
This difference is caused by the $\chi^{12}$ term in Eq. (\ref{tri-t2-in}).
Provided again that the q-dependences of $\chi^{ij}({\bf q})$ around ${\bf Q}$ 
are weaker than that of $F_{\bf q}$,  
$T_{\rm 2M}^{-2}$ at 0$^{\circ}$ and 90$^{\circ}$ can be rewritten as \cite{goto98,goto99a},
\begin{eqnarray}
 [T_{\rm 2M}(0^{\circ})]^{-2} & \propto & \{\chi^{11}({\bf Q})+4\epsilon \chi^{12}({\bf Q})\}^2
\cdot \sum_q \{F_{\bf q}(0^{\circ})\}^2, \nonumber\\
\label{tri-t2-0}
 [T_{\rm 2M}(90^{\circ})]^{-2} &\propto& \{ \chi^{11}({\bf Q}) \}^2 \cdot \sum_q \{F_{\bf q}(90^{\circ})\}^2. 
 \label{tri-t2-90}
\end{eqnarray}
Here, we define the ratio $\rho$ by,
\begin{equation}
 \rho \equiv \frac{[T_{\rm 2M}(0^{\circ})]^{-2}}{[T_{\rm 2M}(90^{\circ})]^{-2}} \cdot {\xi}^{-4}
= \left [ 1+ 4 \epsilon \frac{\chi^{12}({\bf Q})}{\chi^{11}({\bf Q})} \right ]^2,
\label{rho}
\end{equation}
which gives,
\begin{equation}
  \chi^{12}({\bf Q})/\chi^{11}({\bf Q}) = (\sqrt{\rho} -1)/4 \epsilon.
 \label{rchi}
\end{equation}

The $T$ dependence of $\chi^{12}({\bf Q})/\chi^{11}({\bf Q})$ calculated from $\rho$ 
is shown in Fig. \ref{temperature}(c), 
where $\xi$ is adjusted so that $\rho=1$ at high $T$. 
One can see that $\chi^{12}({\bf Q})/\chi^{11}({\bf Q})$ rapidly increases 
in the pseudogap $T$ region, indicating that $\chi^{12}({\bf Q})$ rapidly grows there.
This is consistent with the SEDOR results \cite{stern95,suter99,statt97}
and the theoretical calculations \cite{monien94,kishine97,goto98,hildebrand99,goto99b}.
Note that $\chi^{12}({\bf Q})/\chi^{11}({\bf Q}) \approx T_{\rm 2S}^{-1}/T_{\rm 2G}^{-1}$ 
where $T_{\rm 2S}$ is the SEDOR decay time between the sites on the different layers. 
From Fig. \ref{temperature}(d), 
we find that, 
\begin{equation}
 \chi^{12}({\bf Q})/\chi^{11}({\bf Q}) = 0.28,
\end{equation}
at $T_{\rm c}$,
while $T_{\rm 2S}^{-1}/T_{\rm 2G}^{-1}$ was reported to be about 0.25 at $T_{\rm c}$ 
in Y$_2$Ba$_4$Cu$_7$O$_{15}$.\cite{suter99}
The consistent explanation for both $T_{\rm 2M}^{-1}$ and $T_{\rm 2S}^{-1}$ indicates 
that they can be interpreted on the same basis of the interlayer spin correlations.

In conclusion, we have investigated the anisotropy of the spin echo decay
in the inner Cu site of the trilayer cuprate 
Hg$_{0.8}$Re$_{0.2}$Ba$_2$Ca$_2$Cu$_3$O$_8$
to obtain the anisotropy of the second moment $T_{\rm 2M}^{-2}$.
Comparison between the data at 0$^{\circ}$ and $90^{\circ}$ shows 
the rapid growth of $\chi^{12}({\bf Q})$ in the pseudogap regime.
Since this is a common effect in multilayer systems, 
we conclude that the opposite $T$ dependences 
between $(T_1T)^{-1}$ and $T_{\rm 2G}^{-2}$ observed
in the pseudogap regime of bilayer systems 
are caused by interlayer spin correlations.
The UCLA part of the work was supported by NSF Grant DMR-0072524.

\bibliography{hg1223la}% Produces the bibliography via BibTeX.

\end{document}